# Graphene nano-ribbon under tension

Zhiping Xu

*Department of Engineering Mechanics, Tsinghua University, Beijing, 100084, China*

The mechanical response of graphene nano-ribbon under tensile loading has been investigated using atomistic simulation. Lattice symmetry dependence of elastic properties are found, which fits prediction from Cauchy-Born rule well. Concurrent brittle and ductile behaviors are observed in the failure process at elastic limit, which dominates at low and high temperature respectively. In addition, the free edges of finite width ribbon help to activate bond-flip events and initialize ductile behavior.

Graphene, a new allotrope of carbon, has been obtained experimentally very recently [1]. The two-dimensional nature of this monolayer crystalline material leads to intriguing physical properties, such as a linear dispersion relation of Dirac fermions, anomalous quantum Hall effects and the absence of localization [1]. Besides, the strong $sp^2$ bonding inside the single atomic layer leads to great Young's modulus and tensile stress [2], which opens new perspectives for structural or functional phase in nano-composites [3, 4] and building blocks in nano-electromechanical devices [5]. Especially, graphene nano-ribbon (GNR), tailored from the 2D graphene lattice with finite width, has been found to possess interesting electronic structures with dependence on its width and edge shapes [6, 7]. To find potential applications as reinforcement agents to strengthen and toughen in composites or as electromechanical devices, the mechanical response of the graphene nano-ribbon under tensile loading should be well understood. To this aim, we performed atomistic simulation with focus on the elastic properties and fracture mechanisms.

In our molecular dynamics simulation, the second-generation reactive empirical bond order potential (REBO) [8] was implemented to describe the inter-atomic interaction. This potential is proven to give quantitatively accurate results for mechanical properties of carbon nano-structures. To avoid the unphysical force enhancement caused by the artificial cut-off function in the potential function, we set the cut-off parameters to be 2.0 Å as suggested [9]. To simulate the tensile process, graphene ribbons of length $L = 20$ nm and width $W = 10$ nm are strained at a rate of 0.005%/fs with both ends constrained to maintain the load.

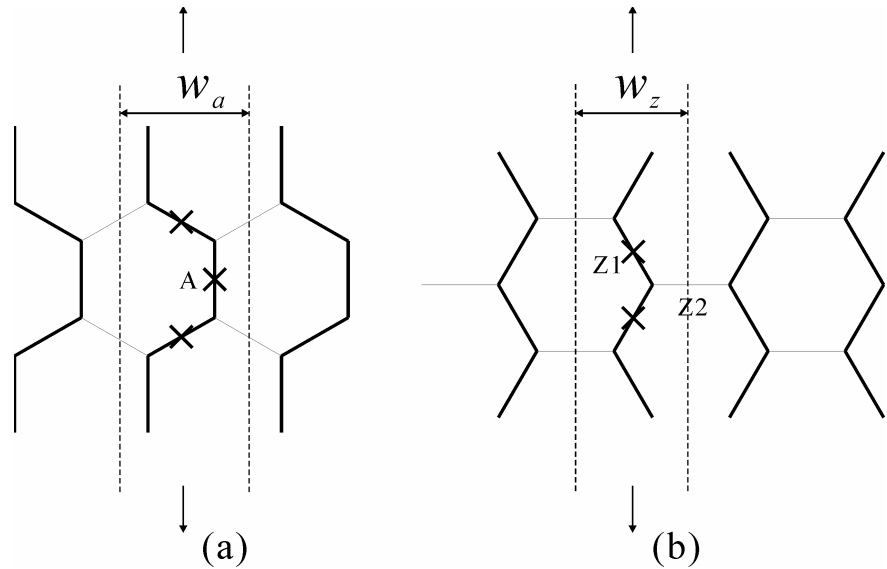

**Figure 1** Graphene nano-ribbon under tension. The bond marked with crossing bears the maximum load; $w_a$ and $w_z$ are the widths of unit area perpendicularly to the ribbon axis. (a) AGNR; (b) ZGNR

The structure of graphene nano-ribbons can be classified into armchair (AGNR), zigzag (ZGNR) and chiral graphene nano-ribbon (CGNR) based on their edge shape [6]. As in the carbon nanotube structures [10], the key feature of the fracture path at low temperature in graphene lattice is the forces distributed in the $sp^2$ bonds. Following the Cauchy-Born rule [9], the bonds A, Z1 and Z2 shown in Figure 1 bears the maximum loads in AGNR and ZGNR. By relating the gross elastic deformation with the individual bond elongation $\delta l_{C-C}$, the brittle-breaking strain of GNR is estimated as [10]

$$\varepsilon_c = 2 \frac{\delta l_{C-C}}{l_{C-C}} \bigg|_c \left[(1-\nu)+(1+\nu)\cos 2\theta\right]^{-1} \qquad (1)$$

where $l_{C-C}$ is the length of carbon-carbon bond and the chiral angle $\theta$ is 0 for AGNR and $\pi/6$ for ZGNR ($\theta = \pi/12$ used in this work) respectively. Thus ZGNR (AGNR) can sustain the largest (smallest) strain before bond breaking in a brittle manner. Using the Poisson's ratio of graphite $\nu = 0.416$ [2], we estimate $\varepsilon_{cAGNR} : \varepsilon_{cCGNR} : \varepsilon_{cZGNR} = 0.65 : 0.71 : 1$, which is consistent with our simulation result $0.67 : 0.74 : 1$ at $T = 300$ K (see Table 1). This consistency indicates the brittle behavior of the fracture process at low temperature, which can also be revealed in the fracture pattern shown in Figure 3. As the Cauchy-Born analysis predicts, the bonds A, Z1 and Z2 under maximum loads have been found to break at first and result in a straight or zigzag crack pattern inside AGNR or ZGNR. We have also calculated the equilibrium elastic properties of GNRs through the stress-strain relation at low strain. The results shown in Table 1 correspond well to the results reported by Reddy et al. [2].

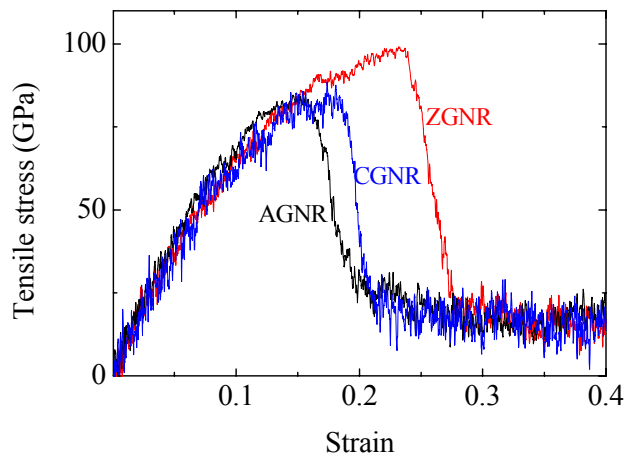

**Figure 2** The stress-strain relation of GNRs with different edge shapes at $T = 300$ K.

Table 1 The mechanical properties of GNRs at 300 K

|  | Young's modulus | Yielding strength | Breaking strain |
| --- | --- | --- | --- |
| AGNR ($\theta = 0$) | 720 GPa | 83 GPa | 0.16 |
| CGNR ($\theta = \pi/12$) | 714 GPa | 85 GPa | 0.175 |
| ZGNR ($\theta = \pi/6$) | 710 GPa | 98 GPa | 0.24 |

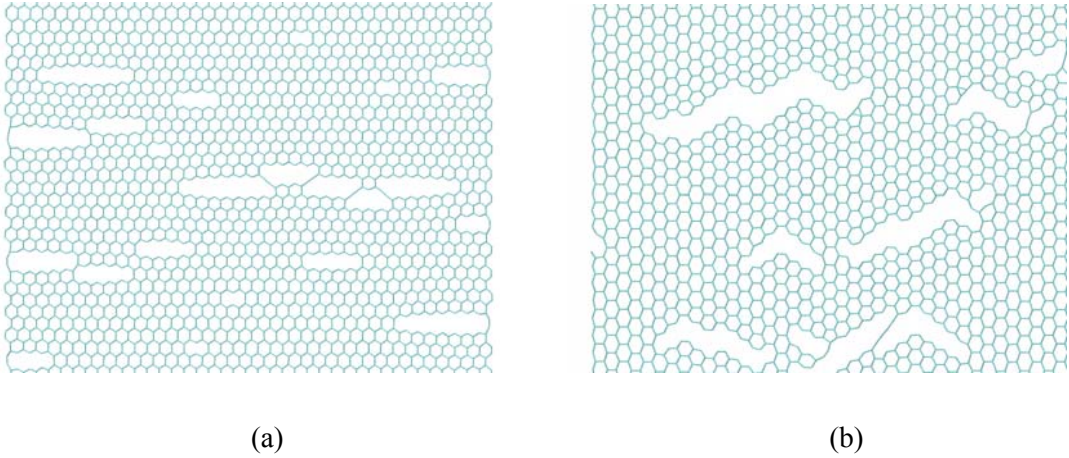

(a)          (b)

**Figure 3** The fracture pattern of AGNR (a) and ZGNR (b) at $T = 300$ K. The ribbon is strained along vertically

    The fracture mechanism of graphene lattice, as revealed by Yakobson et al.[10] in their studies on carbon nanotubes, is dominated by the co-existing ductile flip and brittle breaking of the carbon bonds. At low temperature, the fluctuation of atomic position is prohibited and the bond breaks at elongation limit, thus the behavior can be described well by (1). However, at high temperature the plastic deformation such as SW bond rotation will play a significant role in the fracture process. In our simulation at $T = 2000$ and 3000 K, we have observed frequent bond-flip events (Figure 4a) initialized at the free edge of graphene. The SW defects formed after the bond flipping will propagate inside the ribbon subsequently, and finally induce an edge-localized fracture pattern (Figure 4b). This elasto-plastic deformation observed includes defects diffusion, coalescence and reconstruction[11] in a rather complicate manner.

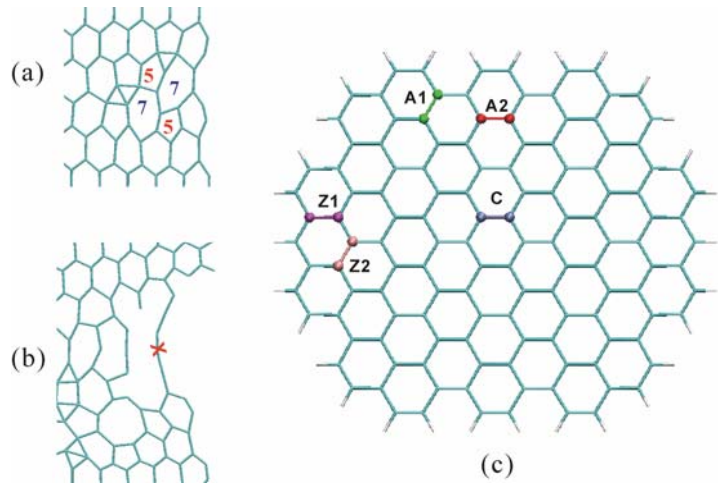

**Figure 4** The Stone-Wales 5/7/7/5 defect (a) and bond breaking event (b) initialized at the free edges of graphene nano-ribbon; different sites of available 5/7 bond-flip events: bonds inside the ribbon (C), bonds close to the armchair edges (A1, A2) and zigzag edges (Z1, Z2).

To explain why the bond-flip events start from the free edge, we present an energy analysis on the SW dislocation at different position in the ribbon through PM3 parameterized semi-empirical calculations [12]. As shown in Figure 4c, in a finite size graphene ribbon, we investigated five representative bond sites: bond in the central ribbon (C) approximating the bond under bulk environment, bonds near the armchair edges (A1, A2) and bonds close to the zigzag edges (Z1, Z2). To calculate the energy of SW defects, we flip these bonds by $\pi/2$ and optimize the structure to a local minimum of potential energy $E_{SW}$. Define as the difference between $E_{SW}$ and the energy of defect-free ribbon $E$, the dislocation energies $E_C$, $E_{A1}$, $E_{A2}$, $E_{Z1}$ and $E_{Z2}$ are calculated and depicted in Figure 5. The dislocation energy is found to be lowered up to 1.46 eV at A1, A2, Z1 and Z2. The decreasing of dislocation energy on the order of 1 eV should be responsible to activation of ductile fracture behavior at the free edges.

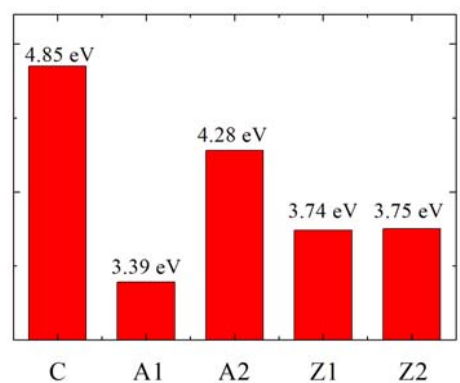

**Figure 5** The Stone-Wales dislocation energy at different bond sites shown in Figure 4(c).

In conclusion, the mechanical properties and fracture mechanisms of nanoscale graphene ribbon have been investigated through atomistic simulations. Brittle bond breakings have been found to dominate at low temperature, and the mechanical response of GNRs in this manner can be described following the Cauchy-Born approach. However, at higher temperature, the plastic deformation assisted by bond-flip events becomes significant. The free edges have been found to lower the SW dislocation energy and will introduce edge initialized ductile fracture process.